\documentclass[twocolumn,showpacs,preprintnumbers,amsmath,amssymb]{revtex4}

\usepackage{graphicx}
\usepackage{dcolumn}
\usepackage{amsmath, bm}
\usepackage{amssymb}

\newcommand{ \bra }[1]{ \langle #1 | }
\newcommand{ \ket }[1]{ | #1 \rangle }

\begin{document}


\title{Production of entanglement in Raman three-level systems using feedback}

\author{R. N. Stevenson}
\author{A. R. R. Carvalho}
\affiliation{Department of Quantum Science, Research School of Physics and Engineering, The Australian National University, ACT 0200, Australia}
\author{J. J. Hope}
\affiliation{Australian Centre for Quantum-Atom Optics, Department of Quantum Science, Research School of Physics and Engineering, The Australian National University, ACT 0200, Australia}

\date{\today}

\begin{abstract}

We examine the theoretical limits of the generation of entanglement in a damped coupled ion-cavity system using jump-based feedback.  Using Raman transitions to produce entanglement between ground states reduces the necessary feedback bandwidth, but does not improve the overall effect of the spontaneous emission on the final entanglement.  We find that the fidelity of the resulting entanglement will be limited by the asymmetries produced by vibrations in the trap, but that the concurrence remains above 0.88 for realistic ion trap sizes. 
\end{abstract}

\maketitle

\section{Introduction}

The production and manipulation of entangled states has been a major feature of quantum information research in the last several years. The efforts in this direction were rewarded with extraordinary experimental advances that led to the realisation of entangled states in a variety of physical systems~\cite{bouwmeester99,rauschenbeutel00,sackett00, panPRL01, roos}, including entanglement involving multiple particles~\cite{Leibfried:2005,Haffner:2005}, and long-lived entangled states~\cite{Roos:2004,Haeffner:2005b}. The latter are achieved either by reducing the experimental imperfections and undesirable interactions with the environment that are responsible for entanglement deterioration, or by encoding entanglement in a decoherence-free subspace~\cite{Zanardi:1997,Lidar:1998}.  A promising approach to deal with the decoherence problem is the use of active quantum feedback control~\cite{belav,wise_milb93,Wiseman:1994,doherty99}. In fact, quantum feedback has been recently proposed and used to improve entanglement production and stability in both continuous~\cite{mancini06,mancini:2006b} and discrete~\cite{Stockton:2004,mancini05,Wang:2005,Carvalho:2007,Carvalho:2008} variable systems. 

Recently, two of us have shown that direct feedback under the appropriate detection strategy leads to the robust production of highly entangled states of two atoms or ions in a cavity~\cite{Carvalho:2007,Carvalho:2008}. Motivated by the perspective of experimental implementations, and the possibility of improving even further the proposed scheme, in this paper we analyse the use of Raman transitions in place of the optical dipole ones considered in~\cite{Carvalho:2007,Carvalho:2008}. A Raman scheme has also been considered by Metz {\it et al.} in a different approach to generate entanglement through the detection of macroscopic quantum jumps~\cite{Metz:2006} without feedback. The entanglement is conditioned on the detection of certain spontaneous emission events that transition the system from a light to a dark state. Moreover, as in~\cite{Carvalho:2007,Carvalho:2008}, entanglement is heralded as prolonged periods of no cavity photon detection indicate that the entangled state has been prepared.

In this paper we will show that the use of feedback in the Metz {\it et al.} setup will allow the desired state to be achieved much quicker, with less time spent in the light state as the transition to the dark state is forced by the feedback, rather than switched through spontaneous emission. We will also show that, although the Raman scheme produces entanglement between metastable levels, which have orders of magnitude smaller decay rates than optical transitions, this does not translate into higher or more stable entanglement in our scheme while feedback is on. In fact, the effect of spontaneous emission from the upper level used for the Raman transition can be detrimental enough to put optical and Raman schemes at the same level in terms of entanglement production. However, the Raman scheme will be shown to significantly reduce the necessary feedback bandwidth.

The paper is organised as follows: In Section~\ref{Sec:Raman} we describe the Raman model and derive the equivalent two-level description after the elimination of the upper level. We establish the equivalence between this model and the one in~\cite{Carvalho:2007} in the absence of spontaneous emission, and present the important effective parameters for the entanglement dynamics. In Section~\ref{Sec:SE} we analyse the effect of spontaneous emission, while in Section~\ref{Sec:Imperfections} we include important experimental limitations as the delocalisation of the ions within the standing wave of the cavity and detection inefficiency. The latter seems to be an important limiting factor for current experimental parameters.

\section{Model}\label{Sec:Raman}

\begin{figure}[h]
\begin{minipage}{18pc}
\includegraphics[width=12pc]{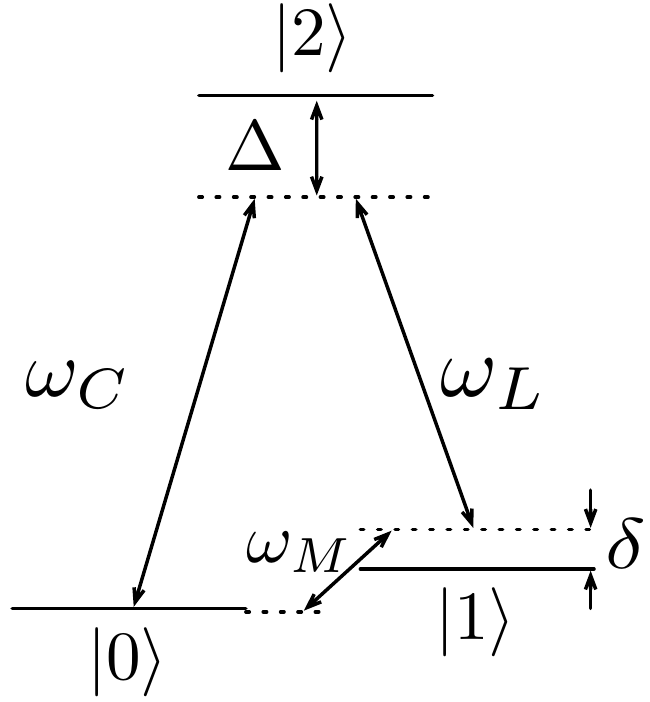}
\caption{\label{levels}
A Raman level scheme. Electrons are excited from level $\ket{0}$ to level $\ket{1}$ by a two photon transition via level $\ket{2}$. The transition with frequency $\omega_L$ is excited by a laser, the transition with frequency $\omega_C$ is excited by the cavity mode and a direct transition between level $\ket{0}$ and $\ket{1}$ transition with frequency $\omega_M$ could be directly excited by microwaves.
}
\end{minipage}\hspace{2pc}%
\end{figure}

The systems consists of a pair of atoms (or ions) with the Raman level configuration shown in Fig.~\ref{levels}. The two target levels are labelled $\ket{0}$ and $\ket{1}$, and have energies $\hbar \omega_0$ and $\hbar \omega_1$ respectively, with the intermediate upper level labelled $\ket{2}$, with energy $\hbar \omega_2$. The particles are coupled to a single mode of an optical cavity which is detuned from the $\ket{0}\rightarrow \ket{2}$ transition by $\Delta$, i.e. $\omega_C = (\omega_2 - \omega_0) - \Delta$. The transition from level $\ket{1}$ to level $\ket{2}$ is pumped using light detuned by $\Delta-\delta$ from resonance, i.e. of frequency $\omega_L = (\omega_2 - \omega_1) - (\Delta-\delta)$, where $\delta << \Delta$, and with a coupling strength $V_L$. This laser pumps the internal levels directly, and does not pump the cavity mode. The transition from level $\ket{0}$ to level $\ket{1}$ is pumped using microwaves detuned from resonance by $\delta$, with coupling strength $V_M$, or equivalently using a different Raman transition, as in \cite{Metz:2007a}. We neglect spontaneous emission for now, but will include it further down. 

Here the particles are assumed to have constant and equal cavity coupling strengths $g$. In section \ref{Sec:Delocalised}, the position of the particles will be allowed to vary in time and therefore the coupling strengths with the cavity mode, $g_a(t)$ and $g_b(t)$, will be time-dependent and different for each particle. 

\begin{figure}[h]
\begin{minipage}{17pc} 
\includegraphics[width=12pc]{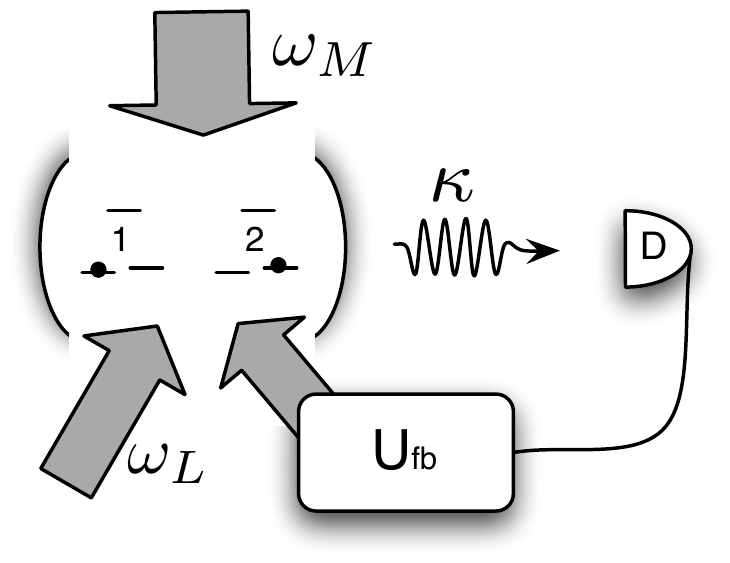}
\caption{\label{modelpic}
A diagram representing the model. The particles are pumped from the side of the cavity. Light leaks from the cavity at a rate $\kappa$ and is detected using a photodetector. When a photon is detected a finite evolution represented by the unitary matrix $\hat{U}_{fb}$ is applied. 
}
\end{minipage} 
\end{figure}

Light that escapes through one of the cavity mirrors is monitored using a photodetector. When a photon is detected, a finite amount of evolution is imposed on the system using control lasers or microwaves. A schematic of this feedback setup is shown in Fig.~\ref{modelpic}.

The master equation for this system, in the interaction picture, using the rotating wave approximation, is
\begin{equation}
	\dot{\hat{\rho}}=-\dfrac{i}{\hbar}\left[\hat{H},\hat{\rho}\right] + \kappa \mathcal{D}\big[\hat{U}\hat{a}\big]\hat{\rho},
\end{equation}
where $\hat{H}$ is the Hamiltonian
\begin{align}
\label{eq:FullH}
	\hat{H}= &-\hbar\delta \hat{A}_{1,1}  + \hbar\Delta \hat{A}_{2,2} \notag \\
	    &+    \dfrac{\hbar V_L}{2}\big( \hat{A}_{1,2}  +  \hat{A}_{2,1} \big) \notag \\
	    &+    \dfrac{\hbar V_M}{2}\big(\hat{A}_{1,0}  +  \hat{A}_{0,1} \big) \notag \\
	    &+    \hbar g\big(\hat{A}_{0,2} \hat{a}^{\dagger}  +  \hat{A}_{2,0} \hat{a}  \big),
\end{align}
with the operators $\hat{A}$ defined as $\hat{A}_{i,j} = \ket{i}_1\bra{j} + \ket{i}_2\bra{j}$, and $\hat{a}$ and $\hat{a}^\dag$ being the annihilation and creation operators of the cavity mode, respectively. $\mathcal{D}$ is the decoherence superoperator
\begin{equation}
\mathcal{D}\big[\hat{c}\big]\hat{\rho} = \hat{c}\hat{\rho}\hat{c}^{\dagger} - \frac{1}{2}\left[\hat{c}^{\dagger}\hat{c}\hat{\rho} + \hat{\rho}\hat{c}^{\dagger}\hat{c} \right],
\end{equation}
and $\hat{U}$ is a unitary matrix describing the effect of the control Hamiltonian $\hat{F}$ applied for a finite time $\delta t$:
\begin{equation}
	\hat{U} = \text{Exp}\left[\dfrac{i \hat{F} \delta t}{\hbar}\right].
\end{equation}
Following the results in~\cite{Carvalho:2007}, we use an asymmetric form of control, acting on either of the particles. We choose $\hat{F} = \alpha (\ket{1}_1\bra{2} + \ket{2}_1\bra{1})$, i.e. $\hat{F} = \alpha (\sigma^-_1 + \sigma^+_1)$ on the first particle, where $\alpha$ represents the strength of the control.

In the regime of large detuning, {\it i.e.}
\begin{equation}
\label{eq:parameters}
	\Delta >> V_L, g, \kappa > V_M,
\end{equation}
we can adiabatically eliminate the upper level and obtain an effective two-level dynamics given by the Hamiltonian
\begin{align}
	\hat{H}_2 = &- \dfrac{\hbar V_L g}{2\Delta} \big(\hat{A}_{0,1} \hat{a}^{\dagger}  +  \hat{A}_{1,0} \hat{a}  \big) \notag\\
	&- \dfrac{\hbar g^2}{\Delta} \hat{a}^{\dagger} \hat{a} \hat{A}_{0,0} + \dfrac{\hbar V_M}{2} \big(\hat{A}_{0,1} +  \hat{A}_{1,0}  \big)\notag\\
	&+ \dfrac{\hbar V_L^2}{4 \Delta} \big(\hat{A}_{1,1} -  \hat{A}_{0,0}  \big) - \hbar \delta \big(\hat{A}_{1,1} -  \hat{A}_{0,0}  \big).
\end{align}
Note that by choosing the second detuning appropriately, $\delta = \frac{V_L^2}{4\Delta}$, the last two terms cancel. 
Now all the terms in the Hamiltonian are on the order of $\frac{1}{\Delta}$ or smaller, so the cavity decay rate $\kappa$ is the fastest rate in the dynamics of the system, and the cavity mode spends most of the time in the zero occupation state. This means the cavity mode can also be adiabatically eliminated in a similar way to~\cite{Wang:2005}. The resulting master equation after eliminating the cavity is 
\begin{align}
\label{ME:TLS}
	\dot{\rho}=  \dfrac{-i V_M}{2}\left[ \left( \hat{A}_{0,1} + \hat{A}_{1,0}\right),\rho\right] 	+ \dfrac{V_L^2 g^2}{\Delta^2\kappa}\mathcal{D}[\hat{U} \hat{A}_{0,1}]\rho.
\end{align}
This is completely equivalent to the feedback equation for a two-level system derived in~\cite{Carvalho:2007}, but with an effective cavity coupling constant of $G_{\text{eff}} = \frac{V_L g}{2 \Delta}$. The system will therefore evolve towards the maximally entangled Bell state $\ket{a_{01}}=\frac{1}{\sqrt{2}}(\ket{01} - \ket{10})$, as shown previously in~\cite{Carvalho:2007, Carvalho:2008}. This jump-based feedback has been shown to be fairly robust against spontaneous emission effects as well as to detection inneficiencies. In the next sections we will investigate the impact of these phemomena on the Raman system, as well as looking at the effect of imperfect trapping of the particles within the cavity mode.

\section{Spontaneous emission effects}\label{Sec:SE} 

Spontaneous emission in a Raman system is significantly different from spontaneous emission in a two-level system. The rate of emission between the two lower levels $\ket{0}$ and $\ket{1}$ is negligible, as the energy difference between them is small. The only significant spontaneous emission is from the upper level $\ket{2}$ to the other two levels. The rates of spontaneous emission are $\gamma_0$ from $\ket{2}$ to $\ket{0}$, and  $\gamma_1$ from $\ket{2}$ to $\ket{1}$, with $\gamma_0 + \gamma_1 = \gamma$. We choose the parameter regime given in Eq.(\ref{eq:parameters}), so that the rate of spontaneous emission is on the order of the other main rates in the system.

The full master equation including the spontaneous emission is
\begin{equation}
	\dot{\hat{\rho}}=-\dfrac{i}{\hbar}\left[\hat{H},\hat{\rho}\right] + \kappa \mathcal{D}\big[\hat{U}\hat{a}\big]\hat{\rho}+ \sum_{i=1,2}\sum_{j = 0,1}\gamma_{j}\bigg(\mathcal{D}\big[\ket{j}_i\bra{2}\big]\hat{\rho}\bigg),
\end{equation}
with the Hamiltonian $\hat{H}$ given by Eq.(\ref{eq:FullH}). One can perform the adiabatic elimination of the upper level and the cavity mode in the same way it was done in Section \ref{Sec:Raman}. In these calculations it is useful to consider a different set of basis states: The symmetric states $\ket{00}$, $\ket{11}$ and $\ket{s_{01}}=\frac{1}{\sqrt{2}}(\ket{01} + \ket{10})$, and the antisymmetric state $\ket{a_{01}}=\frac{1}{\sqrt{2}}(\ket{01} - \ket{10})$. These states are chosen because the Hamiltonian and cavity emission terms are symmetric in the two ions, so they don't couple between the symmetric and antisymmetric subspaces. We can write the final master equation as
\begin{eqnarray}
	\dot{\hat{\rho}}=-\dfrac{i}{\hbar}\left[\hat{H}_{red},\hat{\rho}\right] + \dfrac{V_L^2 g^2}{\Delta^2\kappa} \mathcal{D}\big[\hat{U}\hat{A}_{0,1}\big]\hat{\rho} \nonumber \\ + \sum_{i=s,a}\sum_{j = 0,1}\bigg(\mathcal{D}\big[R_{i,j}\big]\hat{\rho}\bigg),
	\label{Eqn:FullME}
\end{eqnarray}
where $\hat{H}_{red}$ is the adiabatically reduced Hamiltonian as in Eq.(\ref{ME:TLS}):
\begin{equation}
	\hat{H}_{red} = \dfrac{\hbar V_M}{2} \left( \hat{A}_{0,1} + \hat{A}_{1,0}\right).
\end{equation}

Note that the decoherence term can be written in many different ways that are equivalent when it comes to solving the full master equation. Note also, however, that different choices of jump operators can be interpreted as different ways to monitor the system in the framework of quantum trajectories~\cite{Carmichael:1993,Molmer:1993}. In Eq.(\ref{Eqn:FullME}), we followed reference~\cite{Metz:2007a} and broke up the spontaneous emission into four jump terms $R_{i,j}$
\begin{align}
\label{Eq:Rjumps}
	R_{0,s} &= \sqrt{\dfrac{\gamma_0 V_L^2}{4 \Delta^2}}\left(\ket{00}\bra{s_{01}}+ \ket{s_{01}}\bra{11}\right)\notag\\
	R_{0,a} &= \sqrt{\dfrac{\gamma_0 V_L^2}{4 \Delta^2}}\left(\ket{00}\bra{a_{01}}- \ket{a_{01}}\bra{11}\right)\notag\\
	R_{1,s} &= \sqrt{\dfrac{\gamma_1 V_L^2}{8 \Delta^2}}\left(\ket{a_{01}}\bra{a_{01}}+ \ket{s_{01}}\bra{s_{01}} + 2 \ket{11}\bra{11}\right)\notag\\
	R_{1,a} &= \sqrt{\dfrac{\gamma_1 V_L^2}{8 \Delta^2}}\left(\ket{s_{01}}\bra{a_{01}}- \ket{a_{01}}\bra{s_{01}}\right).
\end{align}
This choice reflects the cooperative effects in the fluorescence~\cite{Beige:1999} if one could monitor the quantum jumps using an electron shelving technique~\cite{Sauter:1986}.
In~\cite{Metz:2006}, Metz et al. considered the evolution of the system conditioned on the observation of cavity photons and the spontaneous emission jumps in Eq.(\ref{Eq:Rjumps}) to show that entanglement can be produced in this system, but with no feedback. 
Their results show that some spontaneous emission events bring the system to an entangled dark state, namely the maximally entangled antisymmetric Bell state $\ket{a_{01}}$, and that the system changes from dark (entangled) to light (unentangled) periods according to the measurement events. It was shown that the system spends approximately a quarter of the time in the dark state.

\begin{figure}[h]
	\includegraphics[width=20pc]{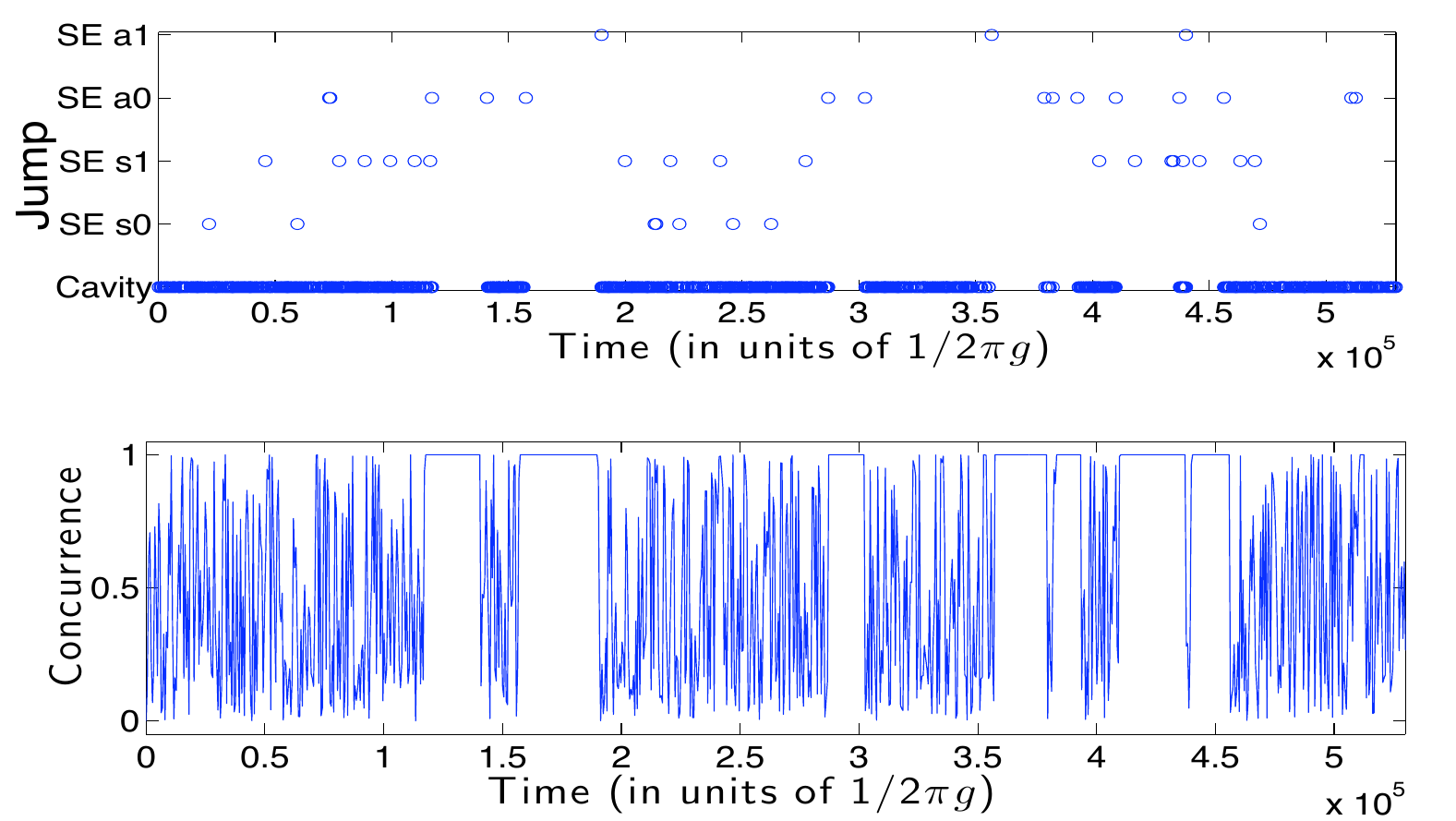}
	\includegraphics[width=20pc]{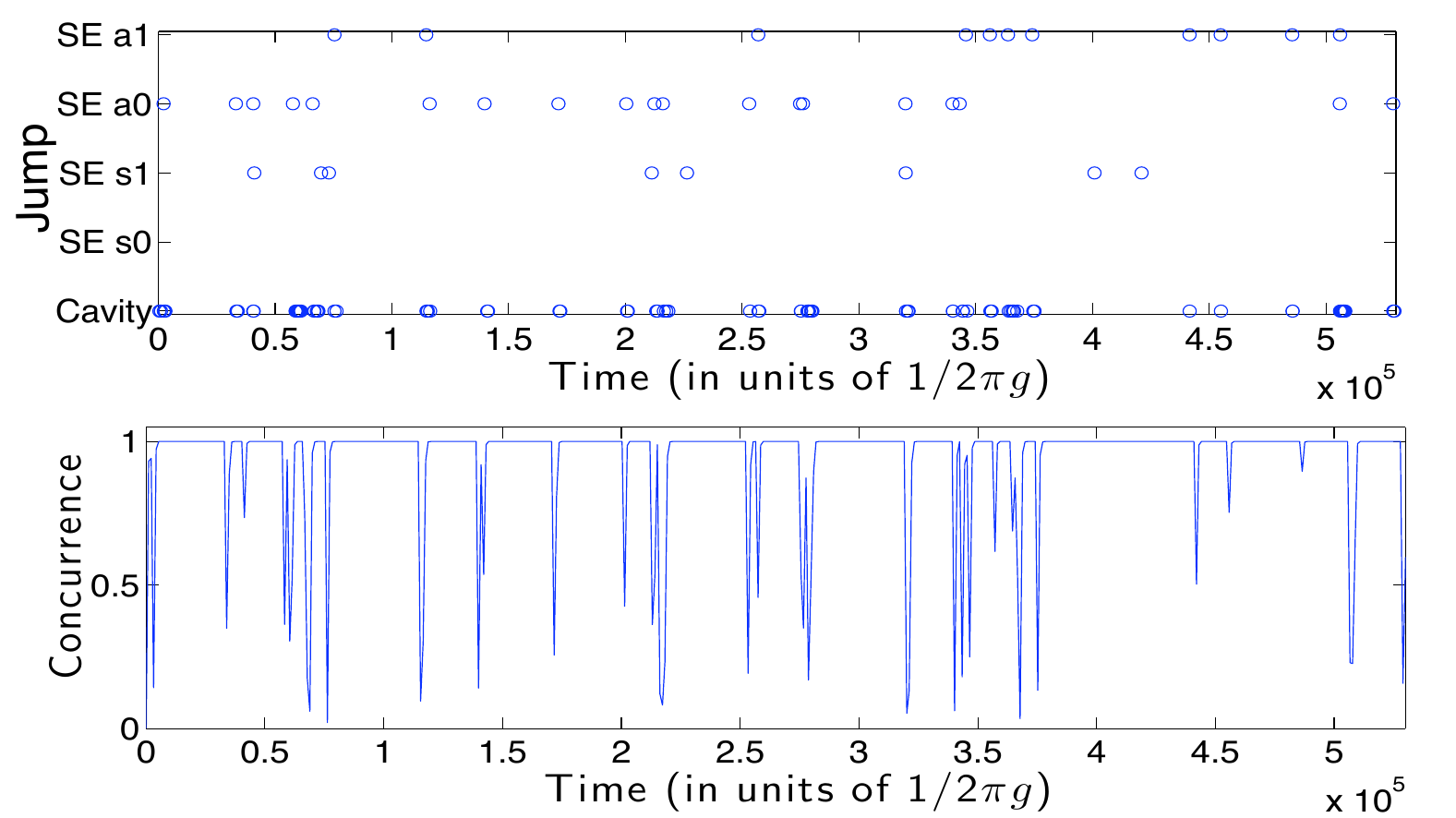}
\caption{\label{Fig:Feedback}
(Color online) Single runs of single path simulations. The top pairs of panels show a run with no feedback, while the bottom pair of panels show the same with feedback. For each of these, the bottom panel shows the concurrence while the top panel shows the occurrence of jumps. The bottom line of jumps is the cavity emission jumps and the rest are spontaneous emission events, with the label corresponting to the jumps in Equations \ref{Eq:Rjumps}. These simulations use the following parameters: $g = \kappa = V_L$, $\gamma = 0.1g$, $V_M = 0.05$ and $\Delta = 50g$, with $\gamma_1 = \gamma_2$ = $g/20$. The feedback Hamiltonian is $\hat{\tilde{F}} = \frac{\hat{F}\delta t}{\hbar} = \frac{\pi}{2}\hat{\sigma}_1$
}
\end{figure}

We have investigated the effect of feedback in this situation and the results are shown in Fig.~\ref{Fig:Feedback}. The entanglement (as measured by concurrence~\cite{Wootters:1998}) and jumps are shown for the cases with (bottom) and without (top) feedback. In the simulations without feedback there is a clear distinction between the dark state, where there are no cavity emissions and the concurrence is 1, and the light state, where there are many cavity emissions and the concurrence oscillates rapidly, much faster than the sampling rate. When feedback is turned on, the duration of the dark periods stays the same but the lengths of the light periods are vastly reduced, and the system spends a greater portion of the time in the desired state. 

\begin{figure}[h]
	\includegraphics[width=20pc]{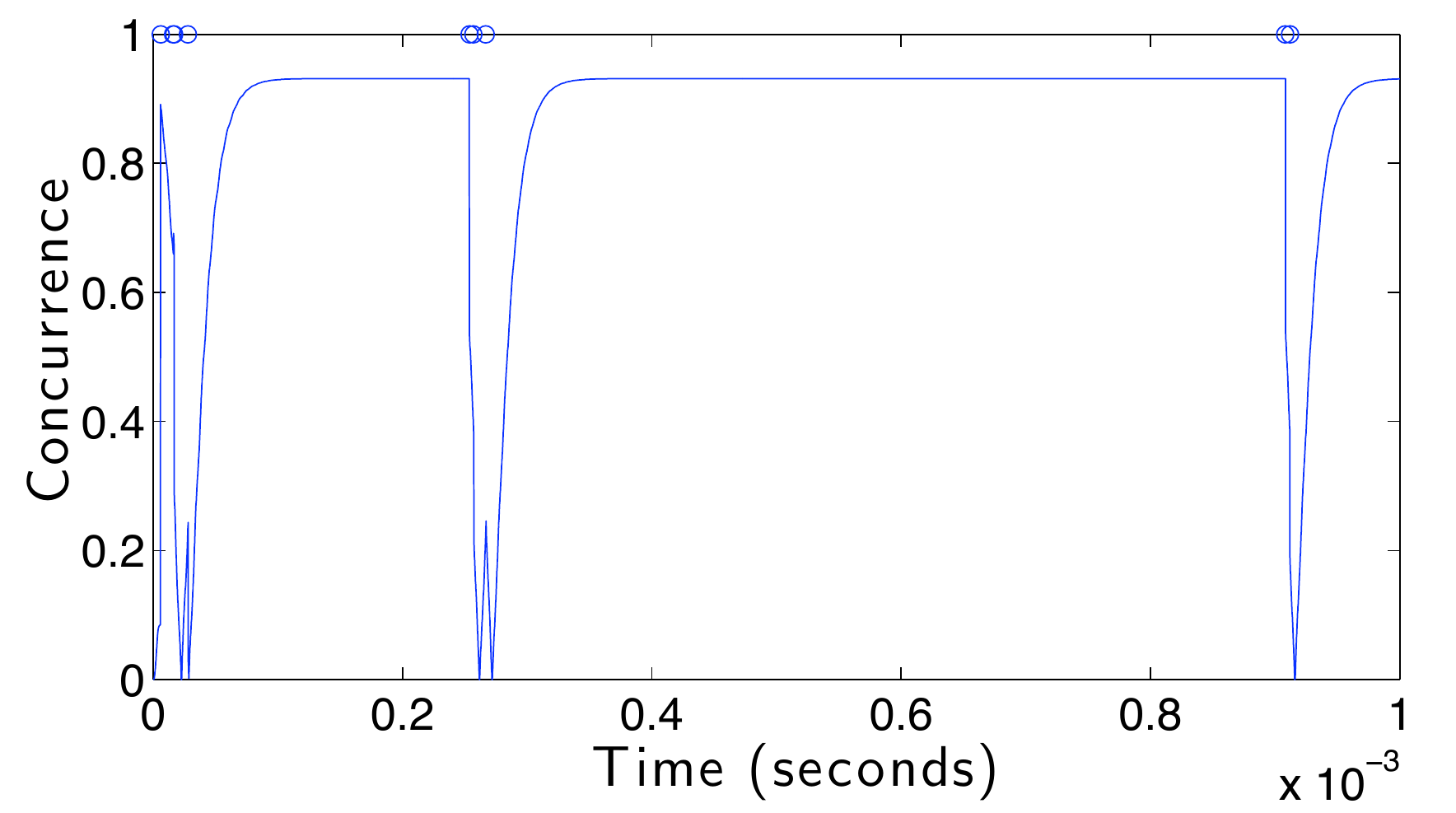}
	\includegraphics[width=20pc]{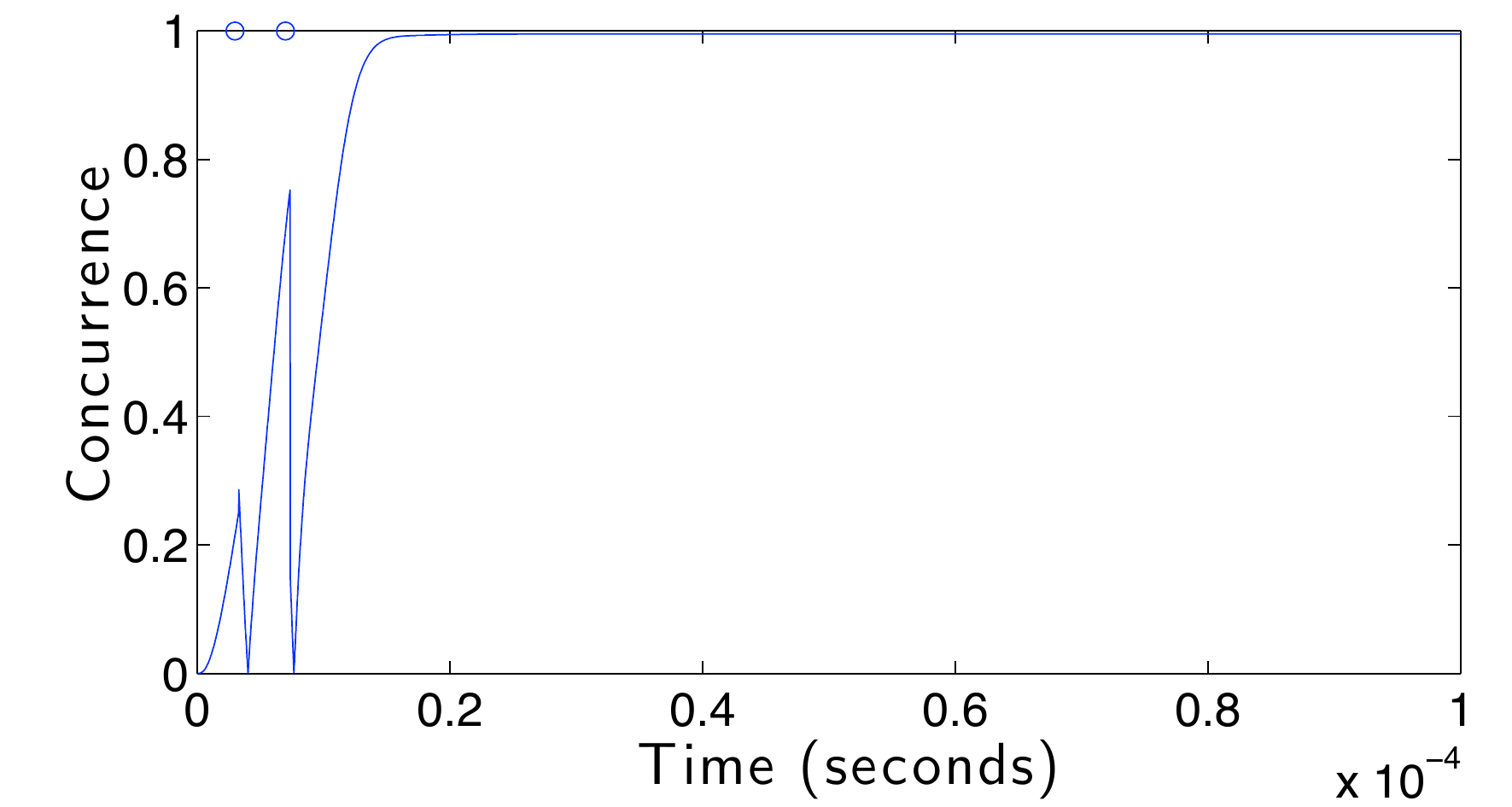}
\caption{\label{Fig:CavityOnly}
(Color online) Simulations of the system with the evolution conditioned on the cavity emission only. These simulations use parameters taken from Russo~\cite{Russo:2008a}(top) and Khudaverdyan \cite{Khudaverdyan:2009}(bottom), with a detuning big enough to be in the adiabatic regime.}
\end{figure}

\begin{figure}[h]
	\includegraphics[width=20pc]{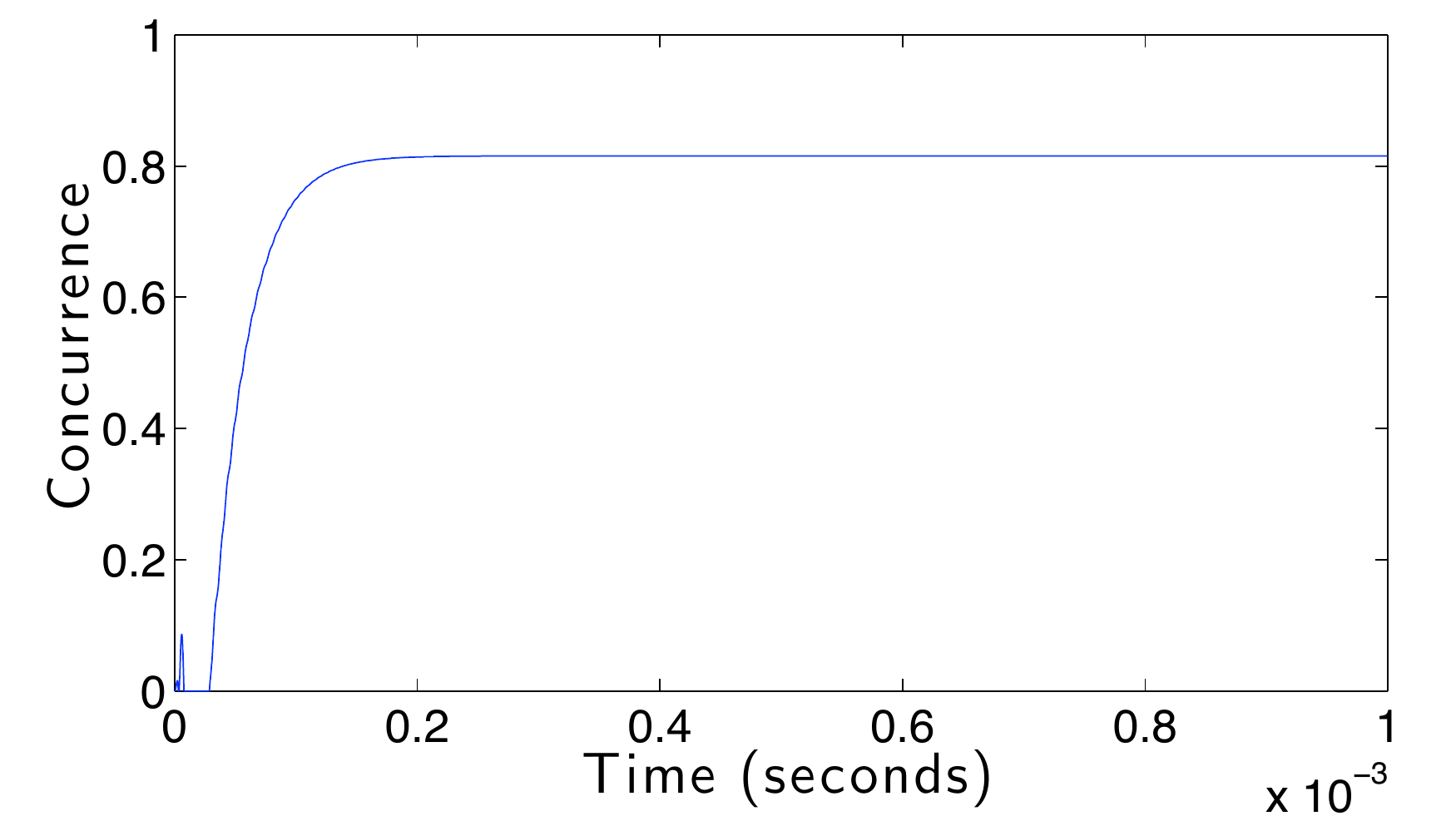}
	\includegraphics[width=20pc]{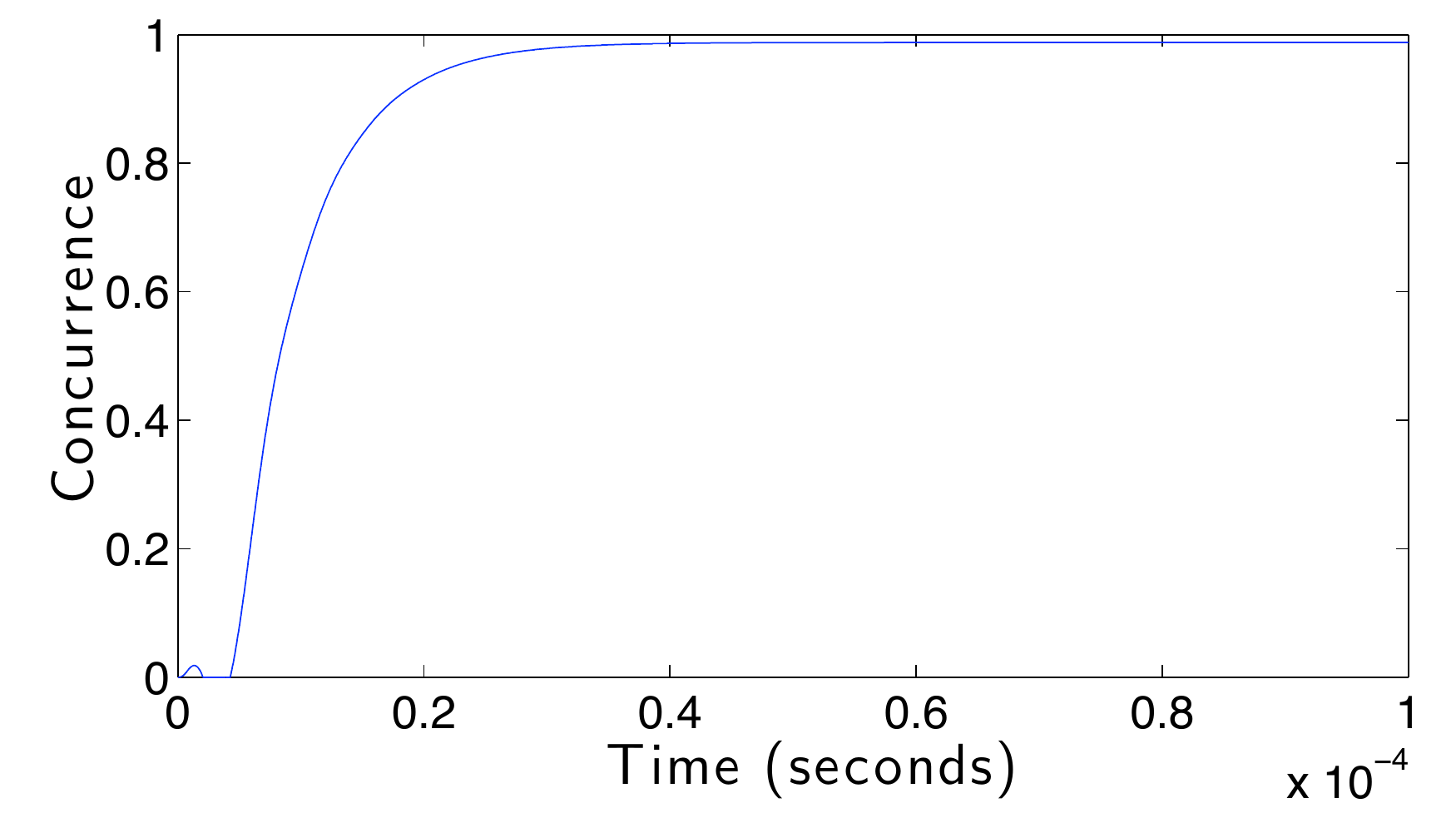}
\caption{\label{Fig:MEfixedG}
(Color online) Entanglement unconditioned on either spontaneous emission or cavity decay.  With no feedback the concurrence is zero, but with feedback concurrence is close to unity. These simulations use parameters taken from Russo~\cite{Russo:2008a}(top) and Khudaverdyan \cite{Khudaverdyan:2009}(bottom), with a detuning big enough to be in the adiabatic regime. 
}
\end{figure}

With the feedback on, the entanglement is robust enough that the system doesn't need to be conditioned on the monitoring of spontaneous emission events.  Monitoring particular kinds of spontaneous emission events would have additional experimental complications and a finite quantum efficiency, which that could degrade the conditional entanglement, so achieving high entanglement without this process is a useful design feature. Figure \ref{Fig:CavityOnly} shows simulations for the system when the evolution is conditioned only on the cavity emissions, and the spontaneous emission is averaged out. We used the experimental parameters for single atoms in a cavity from~\cite{Russo:2008a} ($\{g, \kappa, \gamma\} = 2 \pi \times \{1.61, 0.054, 11.1\} $MHz) (top plot), and from~\cite{Khudaverdyan:2009} ($\{g, \kappa, \gamma\} = 2 \pi \times \{10, 0.4, 2.6\} $MHz) (bottom plot). The detuning $\Delta$ was chosen to be large enough that the parameters were within the two adiabatic regimes described above, with $\Delta = 6$GHz (top) and $\Delta = 0.5$GHz (bottom). These figures are similar to those conditioned on both the cavity emissions and the spontaneous emissions, however the steady state is no longer maximally entangled. If one averages out also the cavity decay, one recovers the solution of the master equation~(\ref{Eqn:FullME}) as shown in Fig.~\ref{Fig:MEfixedG}. 

\subsection{Raman versus Optical scheme}
As can be seen from the results of the previous section, the important factor that determines the amount of entanglement is the ratio between the time the system spends in the maximally entangled state before it decays through spontaneous emission to the time the feedback takes to restore the entangled state after it has decayed. 

The rate at which spontaneous emission occurs is proportional to the population of level $\ket{2}$. The population of this level can be calculated as a function of the populations of the other two levels $\ket{0}$ and $\ket{1}$ as part of the adiabatic elimination process. This leads to an effective spontaneous emission rate of
\begin{equation}
    \gamma_{\rm eff} = \dfrac{V_L \gamma}{4\Delta^2}
\end{equation}
that can be adjusted by changing the detunig of the pump laser and cavity mode. 

To see whether the reduced effective spontaneous emission rate due to the Raman transition is useful, we compare this ratio to the equivalent ratio in a two-level system~\cite{Carvalho:2007,Carvalho:2008}. The rate of feedback in the two-level system is proportional to $\frac{g^2}{\kappa}$ so the entanglement is maximised when $\frac{g^2}{\gamma \kappa}$, which is the single atom cooperativity, is maximised. Taking the equivalent ratio for the Raman scheme, one has, as the new figure of merit,
\begin{equation}
    \frac{G_{\rm eff}^2}{\kappa \gamma_{\rm eff}}=\frac{(\frac{V_L g}{2\Delta})^2}{\frac{V_L^2 \gamma}{4\Delta^2}\kappa} = \frac{g^2}{\gamma \kappa}.
\end{equation}
Unfortunately this shows that, although the rate of spontaneous emission is decreased as the detuning increases, the rate of feedback decreases by the same factor, so that the Raman scheme does not offer advantages over the optical two-level scheme in terms of maximum entanglement produced. It can, however, present some practical advantages. In the optical scheme, the rate at which jumps, and hence control, occur is fixed by the atom-cavity coupling and the cavity decay rate. If this jump rate is too low, entanglement takes longer to build, while if it is too fast, one could reach limitations in the bandwidth of either the photo-detector or the electronics controlling the feedback pulses. On the other hand, the feedback rate in the Raman scheme can be changed by the detuning or laser power, allowing full control over the speed of the process.

\section{Experimental limitations}\label{Sec:Imperfections}

\subsection{Delocalised Particles in a Cavity Standing Wave}\label{Sec:Delocalised}

Another issue with these schemes involves the coupling of the atoms to the optical mode of the cavity.  These models require the coupling strengths of the two atoms to be equal, as this results in the antisymmetric state being a dark state of the system.  The atoms are coupled to standing waves in the cavity, so the coupling strength is proportional to the amplitude of the standing wave (per photon) and as such, varies from maximum to zero in one quarter of a wavelength, typically a couple of hundred nanometres. The field strength varies more gradually in the transverse direction, with the waist of the beam usually being on the order of micrometres. 
Due to their strong interaction with electric fields, ions are a prime candidate for tight trapping, and current ion trapping techniques are able to trap an ion to a little less than an optical wavelength. In \cite{Russo:2008}, $^{40}Ca^+$ ions are trapped to within 70nm. This is not a lot smaller than the distance between two nodes of the standing wave, so the change in coupling strength as the ion moves in the trap will be significant. A similar problem would be present in experiments with neutral atoms.

\begin{figure}[h]
\includegraphics[width=20pc]{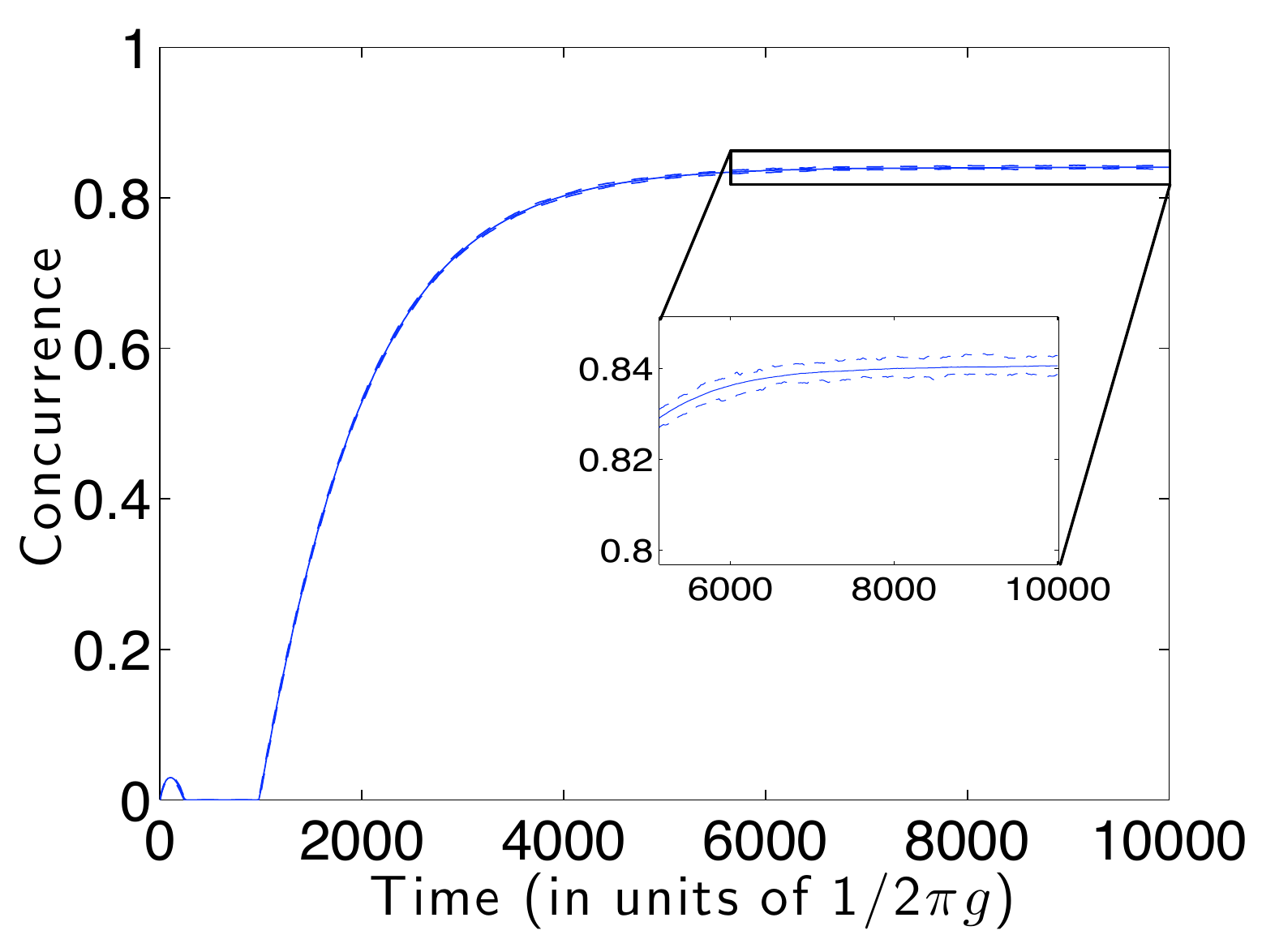}
\caption{\label{Fig:Conc70nm}
(Color online) The concurrence of the two atom system with the atoms trapped in a Gaussian distribution with standard deviation of 0.08 $\lambda$, which approximately corresponds to the trapping of an ion in Reference \cite{Russo:2008}: a 70nm trap within a standing wave with wavelength $\lambda = 866$nm. The solid line shows the average concurrence over 100 runs, with the inset showing the maximum and minimum concurrence over these 100 runs with dashed lines.
}
\end{figure}

The reduced master equation when the coupling constants of the two atoms are allowed to vary is (without spontaneous emission for now),
\begin{align}
	\dot{\hat{\rho}}&=  \dfrac{\hbar V_M}{2}\left[ \left(\hat{A}_{0,1} + \hat{A}_{1,0}\right),\hat{\rho}\right]\notag\\
	&+ \dfrac{V_L^2}{\Delta^2 \kappa}\mathcal{D}[\hat{U}_{fb} (g_{1}(t) \hat{\sigma}_1^- + g_{2}(t) \hat{\sigma}_2^-]\hat{\rho}, \label{Eqn:MEDelocalised}
\end{align}
where $g_i(t)$ are the effective cavity coupling constant for each atom. They vary as $g_i(t) = g_\text{max} cos(\frac{2 \pi x_i(t)}{\lambda})$, with $x_i(t)$ being a Gaussian random position centred at zero with standard deviation of the trap. This corresponds to the centre of the distribution being at an antinode of the standing wave. Figure \ref{Fig:Conc70nm} shows the simulation of Eq.(\ref{Eqn:MEDelocalised}) with the random position re-rolled at each time step. 

\begin{figure}[h]
\includegraphics[width=20pc]{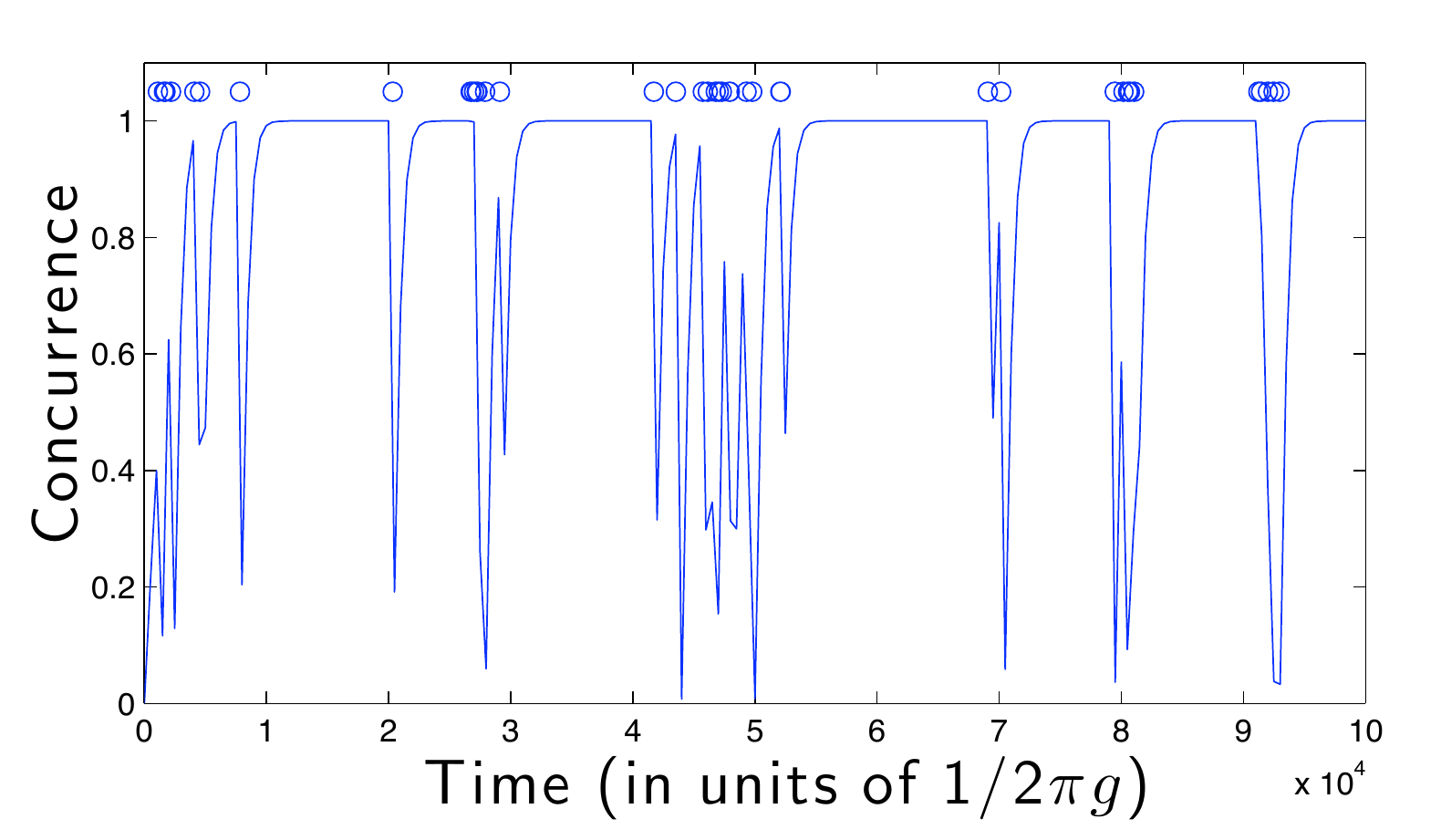}
\caption{\label{Fig:ConcnoSE}
(Color online) The concurrence and cavity emissions for a system with no spontaneous emission and a trap size of 0.2 wavelengths. The solid line shows the concurrence, while the circles show the times when photons leaked out from the cavity.
}
\end{figure}

The reduction in concurrence observed in Fig. \ref{Fig:Conc70nm} can be better understood in terms of the conditional dynamics based on the detection of photons leaving the cavity. As discussed in \cite{Carvalho:2008}, the state $\frac{1}{\sqrt{2}}(\ket{eg} - \ket{ge})$ is the steady state of the system when there is no spontaneous emission and the two coupling strengths are constant.
When the coupling strengths of the two atoms differ, asymmetric elements are introduced into the cavity emission dynamics. This stops the antisymmetric state being a dark state, and the state can now lose excitations through coupling with the environment (decay of the cavity mode), and in this way the concurrence drops. This cavity emission projects the ions into the light, unentangled state. This can be seen in Fig.~\ref{Fig:ConcnoSE}, where there is no spontaneous emission, but there are cavity emission events even when the system is in the maximally entangled anti-symmetric state.

The mean time before a cavity emission from the dark state $\ket{a_{01}}$ is the inverse of the rate of the decay:

\begin{align}
	T_\text{dark}^{-1} &= \dfrac{V_L^2}{\kappa \Delta} \bra{a_{01}} \left( g_{1} \sigma^+_1 + g_{2} \sigma^+_2 \right)\times\notag\\
	& \left( g_{1} \sigma^-_1 + g_{2} \sigma^-_2 \right) \ket{a_{01}}\notag\\
	T_\text{dark} &= \dfrac{2 \kappa}{g_1^2 - g_2^2}.
\end{align}

This is the mean time if the cavity coupling rates are kept constant. As the coupling constants vary, the rate of decay also varies. For the ions to remain in a highly entangled state, this timescale must be much longer than the timescale over which the feedback drives the ions to the dark state.

\subsection{Non-unity cavity emission}

The effect of having photodetectors without perfect efficiency is the same in the Raman system as it is in a two level system. When the photodetector has an efficiency $\eta$, the master equation (\ref{Eqn:FullME}) is transformed into~\cite{Carvalho:2008}

\begin{align}
	\dot{\hat{\rho}}&=  \dfrac{-i}{\hbar}\left[\hat{H},\hat{\rho}\right]\notag\\
	&+ \eta\dfrac{V_L^2}{\kappa \Delta^2}\mathcal{D}[\hat{U}_{fb} (g_{1}(t) \hat{\sigma}_1^- + g_{2}(t) \hat{\sigma}_2^-]\hat{\rho}\notag\\
	&+ \left(1-\eta\right)\dfrac{V_L^2}{\kappa \Delta^2}\mathcal{D}[(g_{1}(t) \hat{\sigma}_1^- + g_{2}(t) \hat{\sigma}_2^-]\hat{\rho}\notag\\
	&+ \sum_{i=s,a}\sum_{j = 0,1}\bigg(\mathcal{D}\big[R_{i,j}\big]\hat{\rho}\bigg). \label{Eqn:MENonUnityDetectors}
\end{align}
This is simply saying that when the photon is detected by the photodetector (the term with the $\eta$ prefactor) the feedback is applied, and when the photon is not detectected (the term with the $(1-\eta)$ prefactor) no feedback is applied. A cavity emission event is essentially a lost opportunity to apply feedback, so the photodetector efficiency is a scaling factor on the rate at which feedback is applied. 

\begin{figure}[h]
\includegraphics[width=20pc]{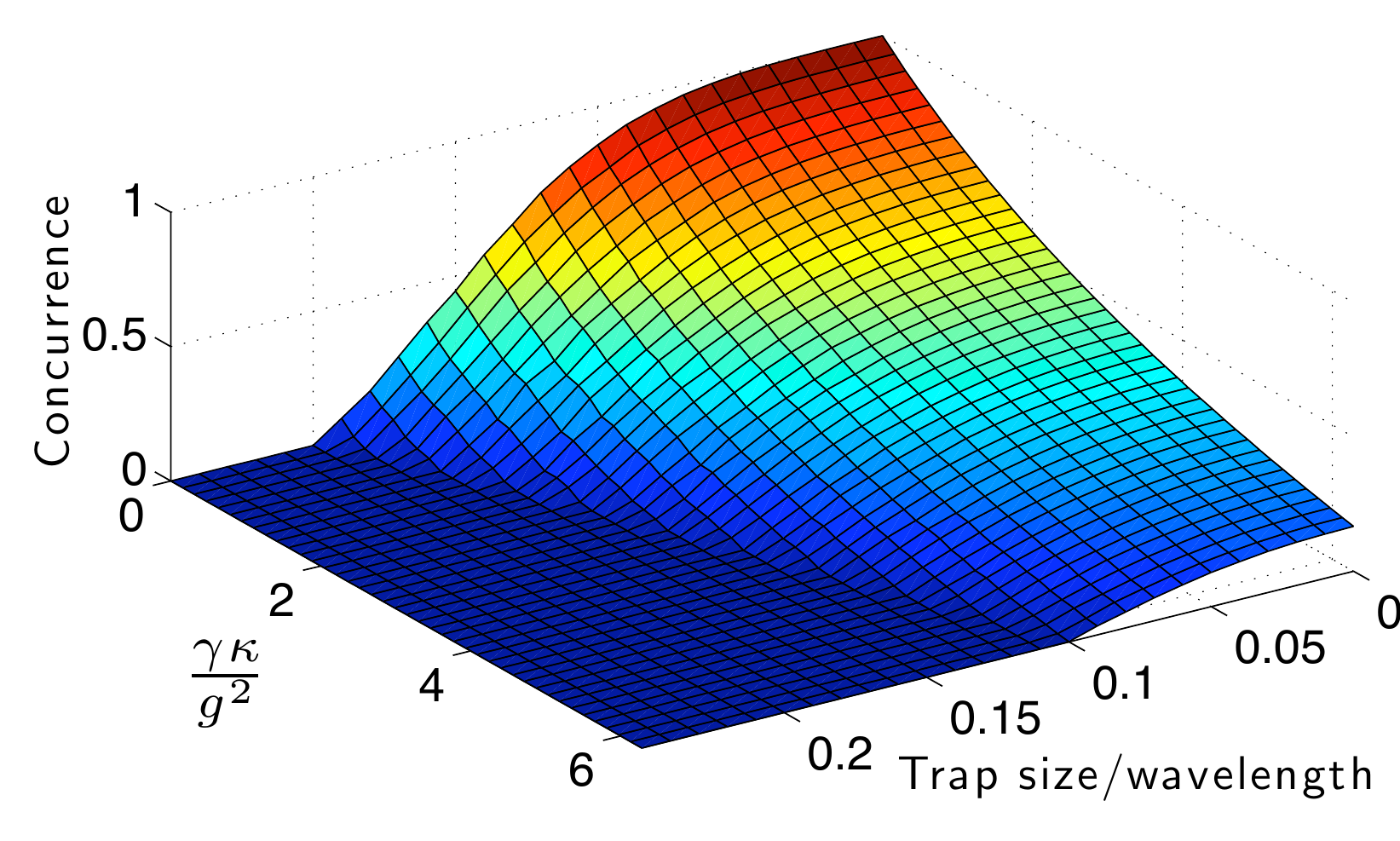}
\includegraphics[width=20pc]{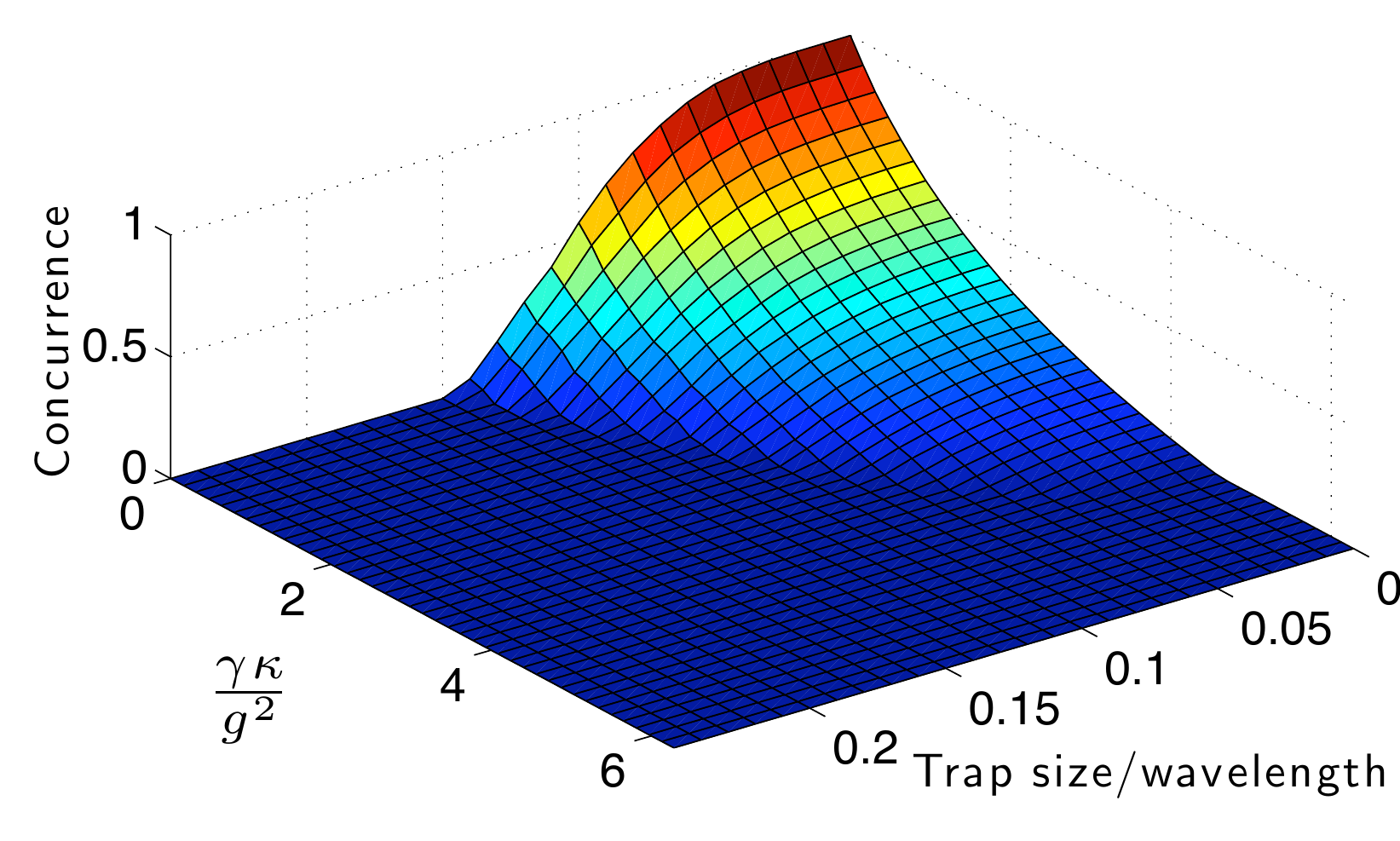}
\caption{\label{Fig:ConcScanggamma}
(Color online) The top shows the steady state concurrence of the two atom system as a function of the size of the trap and the spontaneous emission rate. The x axis is a measure of the standard deviation of the particles as a fraction of the cavity mode wavelength $\lambda$. The y axis is the total spontaneous emission $\gamma$ rate as a fraction of $g$. The z axis is the steady state concurrence. The bottom figure shows the same with a detector efficiency of 50\%. 
}
\end{figure}

A summary of the effects of detection inefficiencies and cavity coupling fluctuations can be seen in Fig. \ref{Fig:ConcScanggamma}. The steady-state concurrence for a range of trap sizes and spontaneous emission rates running the full master equation simulations is shown. For reasonable parameters, $\gamma = 0.1 g$ and with ions trapped within .08 of a wavelength, the concurrence is above 0.88. As the trap size increases past 0.1 wavelengths the concurrence rapidly drops.

\section{Conclusions}

In this paper we have shown that a feedback control scheme to generate entanglement using Raman three-level atoms is mostly equivalent to a system using two-level optical transitions. In the regimes where cavity mode and upper level can be adiabatically eliminated, the final entanglement depends on the cooperativity parameter that is the same for Raman or optical schemes. While this equivalence is true while feedback is on, the Raman system has two main advantages.  First, the necessary feedback bandwidth can be experimentally controlled, rather than having to operate at the lifetimes of optical transitions.  Second, the Raman scheme has a higher entanglement lifetime if feedback is turned off. 

Feedback prepares unconditioned entangled states with concurrence greater than 0.88 even when taking into account a realistic delocalisation of the atoms within the cavity mode. Entanglement can be even higher, and heralded, if the system is conditioned on the detection of photons leaving the cavity. 
Reliably produced pairs of entangled states could be used to make entangled states of more particles \cite{Metz:2007}, with an ultimate aim of producing large cluster states for measurement based quantum computing. This opens the perspective to produce multipartite entangled states directly through a similar feedback setup, by choosing appropriate measurement and controls.

\bibliography{Papers}

\end{document}